\documentstyle[aps,prb,epsf]{revtex}
\draft
\begin{document}
\def\SNG{{\em Physical Review Style and Notation Guide}}
\def\LUG {{\em \LaTeX{} User's Guide \& Reference Manual}}
\def\btt#1{{\tt$\backslash$\string#1}}%
\def\REVTeX{REV\TeX}
\def\AmS{{\protect\the\textfont2
        A\kern-.1667em\lower.5ex\hbox{M}\kern-.125emS}}
\def\AmSLaTeX{\AmS-\LaTeX}
\def\BibTeX{\rm B{\sc ib}\TeX}
\twocolumn[\hsize\textwidth\columnwidth\hsize\csname@twocolumnfalse%
\endcsname
\title{Multiband Electronic Raman Scattering in Bilayer Superconductors}
\author{T. P. Devereaux$^{1}$, A. Virosztek$^{3,2}$, and A. 
Zawadowski$^{2,3}$}
\address{$^{1}$Department of Physics, University of California, Davis, CA 95616}
\address{$^{2}$Institute of Physics and Research Group of the Hungarian
Academy of Sciences, Technical University of Budapest,
H-1521 Budapest, Hungary}
\address{$^{3}$Research Institute for Solid State Physics, P.O. Box 49,
H-1525 Budapest, Hungary}
\date{\today}
\maketitle
\begin{abstract}
A theory of electronic Raman scattering in the presence of several
energy bands crossing the Fermi surface is developed. The contributions
to the light scattering cross section are calculated for each band
and it is shown that the cross section can be written in terms of
the sum of the single band contributions and a mixing term which
only contributes to the fully symmetric channels $(A_{1g})$. Particular
emphasis is placed on screening in bilayer superconductors. Since any
charge fluctuation with long range character in real space is
screened by the Coulomb interaction, the relevant fluctuations in a
single layer case are induced between different parts of the Fermi
surface. In a single band $d$-wave superconductor the scattering at
energy transfer twice the maximum gap $\Delta_{max}$ is dominated by
those parts of the Fermi surface where $\Delta^2({\bf k})$ is largest.
As a consequence, the fully symmetric ($A_{1g}$) scattering is screened.
In the case of a bilayer superconductor however, the charge transfer is
possible between layers inside the unit cell. Therefore
a formalism is considered
which is valid for general band structure, superconducting energy gaps, 
and inter-layer hopping matrix elements. The spectra is calculated for
La 2:1:4 and Y 1:2:3 as representative single and bilayer superconductors. 
The mixing term is found to be negligible and thus the response is well
approximated by the sum of the contributions from the individual bands. 
The theory imposes
strong constraints on both the magnitude and symmetry of the energy gap 
for the bilayer cuprates, and indicates that a nearly identical energy gap
of $d_{x^{2}-y^{2}}$ symmetry provides a best fit to the data. However,
the $A_{1g}$ part of the spectrum depends sensitively on many parameters.
\end{abstract}
\vskip 1cm
]
\section{introduction}

In the recent few years Raman scattering has been proved to be a very
powerful tool to study the anisotropy of the superconducting energy gap
in high-T$_{c}$ superconductors\cite{one,carbotte,irwin,candk}. Even if 
Raman scattering (in clean materials) can give
information only about the absolute value of the superconducting gap,
the spectra clearly indicate the extremal points of the gap as the
maxima and minima which show up as singularities in the electronic
density of states. The spectra show that the 
electronic part has essential density of states below the maxima associated with the
maximum of the energy gap. On the other hand, the data clearly
shows that there is only one sharp maximum in the spectra, indicating
that the absolute value of the maximum and minimum of the gap are
very close\cite{exp,slakey,alt}. Thus the gap smoothly interpolates between $\Delta_{max}$
and $-\Delta_{max}$ and no other singular part exists. Such a gap
occurs in $d-$wave superconductors. 
One can clearly state that if the gap contained mixtures of different
symmetries, the $d-$wave part at least must be the dominant one\cite{one}.

The experiments carried out with different polarization orientations
pick up the contributions to the light scattering on different
parts of the Fermi surface. For example, incident and scattered
light polarizations ${\bf e}^{I,S}$ aligned
along $\hat x+\hat y, \hat x+
\hat y$ or $\hat x +\hat y, \hat x - \hat y$, select then the
$A_{1g}+B_{2g}$ or $B_{1g}$ part of the spectra is measured, where
the $x,y$ coordinate system is locked to the CuO$_{2}$ planes. Measuring
the $B_{1g}$ spectra provides information mainly concerning
the light scattering in the neighborhood of the $k_{x}$ and $k_{y}$ axes,
$B_{2g}$ spectra probes mainly along the diagonals, and $A_{1g}$ is a
weighted average over the entire Brillouin zone. The sharpest
spectra are observed in the $B_{1g}$ channel and thus the maxima of
the absolute value of the energy gap must lie along the axes,
supporting a gap of $d_{x^{2}-y^{2}}$ symmetry. The recent experiments
carried out in different hole doped
high-T$_{c}$ materials provide almost
identical spectra showing very characteristic differences between
the various polarization orientations\cite{exp}.

In forming the Raman spectra, long-range Coulomb screening plays
a very important role. Namely, those parts of the spectra which are
coupled to the Coulomb fields are screened and do not appear in the 
limit of small momentum transfer $q \rightarrow 0$, as is always
the case. The long-range Coulomb forces play an important role if
there is charge transfer between unit cells at large distances. Thus
only the charges produced by charge transfers inside the unit cell
determine the measurable spectra. This transfer is between the different
atoms in the cell or the redistribution of the electrons between
different parts of the Fermi surface or between different Fermi surfaces.
In this way, only the $A_{1g}$ part of the spectra can be coupled to the
Coulomb forces.  This argument is based on the assumption that only one
Fermi surface is relevant, which is very reasonable for materials with
a single CuO$_{2}$ plane in the unit cell. Recently Krantz and Cardona
\cite{candk} have raised the relevant question how a double-sheeted Fermi
surface, as occurs in materials with more than one CuO$_{2}$ planes in
the unit cell, changes the above argument. 

The general idea of charge transfers between different regions of 
${\bf k}$ space with energies nearby the Fermi level is very similar
to the case of semiconductors, where the electrons or holes can be
redistributed among several valleys of the conduction bands or the
different parts of the valence bands\cite{semin}.

Except for scattering in the $A_{1g}$ channel.  
the presence of the double layer does not affect essentially the
theory of Raman scattering as the Coulomb screening
does not play a role. In the $A_{1g}$ symmetry, however, additional
effects may occur as expected by Ref.\cite{candk}. Therefore, the
$A_{1g}$ channel will be examined in great detail in the present paper.

The main question is whether or not in the double layer compound a singularity
can occur at 2$\Delta_{max}$ in the $A_{1g}$ symmetry, similar to the
one which occurs in the $B_{1g}$ case. In the single layer compounds the
calculations show that the peak with $A_{1g}$ symmetry is far from being
singular, with a position ranging from somewhat near $\Delta_{max}$
\cite{one,carbotte} to slightly below $2\Delta_{max}$\cite{candk}.

The consideration will be focused on the energy range $\omega \sim
2\Delta_{max}$, thus nearby the singular contribution. If there is a 
singularity at $2\Delta_{max}$, then the region nearby the axes
dominates the spectrum and from the point of view of screening
the other parts can be ignored as their relative weight disappears
as $\omega \rightarrow 2\Delta_{max}$. In Fig. 1, those regions
of are shown, which is also where the absolute value of the gap
is maximal.  There, the Fermi surface of the two layers are shown.
Of course, in a more appropriate discussion followed in the present
paper, the even and odd (bonding and anti-bonding) combinations of the
wave functions of the layers must be introduced. In order to first get 
an insight the labels of the CuO$_{2}$ layers is used. Considering the
electronic density, the singular regions of the Fermi surface are
indicated by small ellipsoids with a strong resemblance to the different
valleys in the case of multivallied semiconductors. The two layers
double the number of these regions. In the case of $B_{1g}$ symmetry
the charge transfer is between the region of the $k_{x}$ and $k_{y}$
axes, respectively, and changes sign going from one region to the other.
In contrast, for $A_{1g}$ symmetry, each region has the same
phase on a particular Fermi surface. Therefore, in the case of the 
single layer material, considering the singular part, the charge
transfer is only between different cells, which is screened. In the 
double layer case, the $A_{1g}$ symmetry allows a charge transfer 
between the two layers of both the same and opposite sign on each
Fermi surface. Using the bonding and anti-bonding wave functions of
the two layers, the odd combination plays an important role.

Considering the interlayer charge transfer process, the question
must be raised how the light coupled to both planes in a similar
way can result in a charge transfer between the symmetrically
displaced layers. That discrepancy can be resolved only by
considering the coupling between the planes which automatically leads
to the introduction of the a bonding and anti-bonding wave functions.
The coupling leads to a split of the two bands which is also a measure of
the possible charge transfer between the planes, as will be discussed in
the present paper. A nearly singular peak at $\omega \sim 2\Delta_{max}$
can occur if the splitting of the superconducting gaps for the electrons
with even and odd wave functions is small. The strength of the peak is
proportional to the square of the difference between the Raman efficiency
$(\gamma_{+}-\gamma_{-})^{2}$, where $\gamma_{\pm}$ correspond to the
coupling of light to the bonding and anti-bonding electrons, respectively.
This difference can be demonstrated in the limit where the incident and
scattered photon energies are small compared to the relevant band
energies. In this limit, $\gamma_{+}-\gamma_{-}$ is proportional to the
splitting of the effective mass in the two bands. It turns out that
considering a realistic description of YBa$_{2}$Cu$_{3}$O$_{7}$ and
Bi$_{2}$Sr$_{2}$CaCu$_{2}$O$_{8}$, the splitting of the mass is less than 
10 percent\cite{photo}. Thus the peak at
$\omega=2\Delta_{max}$ can not have a larger relative intensity 
then $10^{-2}$.  If that argument holds, the $A_{1g}$ spectra must be
very similar to those calculated from the two separate uncoupled layers
and the effect of the coupling is very likely to be beyond the 
observable range.

In our paper, we consider
the most representative of the single and double layer compounds La 2:1:4 
and Y 1:2:3, respectively, while we will only briefly mention
the multilayer compounds, such as Tl 2:2:3:2, in the Appendix
since the main physical ideas can be best illustrated by contrasting 
the single and double sheet compounds.

\section{model and calculations}

The simple model which contains the physics of bilayer systems is
given by
\begin{eqnarray}
H=\sum_{{\bf k},\alpha}\epsilon_{\alpha}({\bf k})c^{\dagger}_{\alpha}({\bf k})
c_{\alpha}({\bf k})\\
+{1\over{2}}
\sum_{{\bf k},\alpha\ne\beta}t_{\alpha,\beta}({\bf k})
c^{\dagger}_{\alpha}({\bf k})c_{\beta}({\bf k})+h.c.,\nonumber
\end{eqnarray}
where ${\bf k}=(k_{x},k_{y})$ is the in-plane momentum,
$\alpha,\beta=1,2$ represent the given plane indices and the sum
over spins is implicitly included. $t_{\alpha,\beta}$ represents the
interplane coupling, through which electrons can hop from one
plane to the other directly or through an intermediate state
such as the chains or Y atoms. $\epsilon_{\alpha}({\bf k})$ are the band
energies for the separate planes in the absence of an interlayer
coupling. 

Due to the mirror plane symmetry of the bilayer,
the above Hamiltonian can be diagonalized by considering even and
odd combinations of electron operators on the two planes. We
introduce the new operators
\begin{equation}
c_{\pm}^{\dagger}({\bf k})={1\over{\sqrt{2}}}[c_{1}^{\dagger}({\bf k})\pm c_{2}^{\dagger}({\bf k})]
\end{equation}
allowing us to rewrite the Hamiltonian in diagonal form
\begin{equation}
H=\sum_{k}[\epsilon_{+}({\bf k})c_{+}^{\dagger}({\bf k})c_{+}({\bf k})
+\epsilon_{-}({\bf k})c_{-}^{\dagger}({\bf k})c_{-}({\bf k})],
\end{equation}
with the bonding and anti-bonding band energies given 
by $\epsilon_{\pm}({\bf k})=
\epsilon({\bf k})\pm t_{\perp}({\bf k})$, respectively,
where for simplicity we
have taken degenerate bands in the absence of
interlayer coupling: $\epsilon_{1}({\bf k})=\epsilon_{2}({\bf k})
=\epsilon({\bf k})$. Also we have defined $t_{\perp}({\bf k})=
t_{1,2}({\bf k})=t_{2,1}({\bf k})$.

The intensity of Raman scattering can be given as a scattering off
an effective charge\cite{kandd}
\begin{equation}
\tilde\rho=\sum_{{\bf k},\alpha,\beta}\gamma_{\alpha,\beta}({\bf k})
c^{\dagger}_{\alpha}({\bf k})c_{\beta}({\bf k}),
\end{equation}
where $\alpha,\beta$ are the plane indices and $\gamma({\bf k})$ is
the Raman scattering amplitude. The effective charge is a nonconserving
quantity in contrast to the real charge.
Here the Raman vertices $\gamma$
are related to the incident and scattering photon
polarization vectors ${\bf e}^{I,S}$, resulting from a coupling of
both the charge 
current and the charge density to the vector potential. 
In general the Raman vertex
depends non-trivially on both the incident and scattered photon
frequencies. In the case of a bilayer system the light can interact
with the electrons on the planes or the atoms in between. The effective
charge can be rewritten in terms of the bonding and anti-bonding
combinations of the bands,
\begin{eqnarray}
\tilde\rho=\sum_{\bf k}\biggl\{\gamma({\bf k})[c^{\dagger}_{+}({\bf k})
c_{+}({\bf k})+
c^{\dagger}_{-}({\bf k})c_{-}({\bf k})]\\
+\gamma_{1,2}({\bf k})
[c^{\dagger}_{+}({\bf k})
c_{+}({\bf k})-c^{\dagger}_{-}({\bf k})c_{-}({\bf k})]\biggr\},\nonumber
\end{eqnarray}
where
\begin{eqnarray}
\gamma({\bf k})=\gamma_{1,1}({\bf k})=\gamma_{2.2}({\bf k})=
{1\over{2}}[\gamma_{+}({\bf k})+\gamma_{-}({\bf k})], \\
\gamma_{1,2}({\bf k})={1\over{2}}[\gamma_{+}({\bf k})-\gamma_{-}(\bf k)].
\nonumber
\end{eqnarray}
This corresponds to a diagonal contribution which allows light to be
scattered by density-like
fluctuations on either plane 1 ($\gamma_{1,1}$) or plane
2 ($\gamma_{2,2}$) and an off-diagonal term ($\gamma_{1,2}$) which allows
light scattering on both planes simultaneously (see Fig. 2). Since in the
limit of ${\bf q} \rightarrow 0$ the long-range intercell fluctuations
are screened, only the intracell fluctuations remain. Therefore only
charge transfer fluctuations between the atoms in the plane and outside
the plane (e.g., $Y$ and the chains), and between the planes can effectively
cause light scattering. 
Since experimentally the interlayer coupling is small, the resulting
vertex $\gamma_{1,2}$ must be smaller than the $+$ combination, labelled
as $\gamma$. Similar considerations have been applied to the
case of multi-valley scattering in the conduction band of semiconductors. 
In the case when the
intervalley scattering does not take place, no light scattering can occur as 
the density fluctuations
are screened completely. Scattering is restored when inequivalent 
valleys are considered or the scattering is between different parts of the
valence band.

According to Abrikosov and Genkin\cite{ag} if the energy of the incident
and scattered frequencies $\omega_{I},\omega_{S}$ are negligible
compared to the relevant energy scale of the band structure, $\gamma_{\pm}$
can be expressed in terms of the curvature of the bands and the incident
and scattered photon polarization vectors ${\bf e}^{I,S}$ as
\begin{equation}
\lim_{\omega_{I},\omega_{S} \rightarrow 0}\gamma_{\pm}({\bf k})
=\sum_{\mu,nu} e^{I}_{\mu}{\partial^{2}\epsilon_{\pm}({\bf k})\over{\partial k_{\mu}
\partial k_{\nu}}}e^{S}_{\nu},
\end{equation}
where terms of the order of $1-\omega_{S}/\omega_{I}$ are dropped.
This expression is valid given that the incoming laser light cannot excite
band-band transitions. This effective mass approximation has been extensively used 
to calculate the Raman spectra in unconventional superconductors. 

However one might question the 
appropriateness of this approximation for the cuprates given
that typical incoming laser frequencies are on the order of 2 eV -  certainly
on the order of 
the relevant electronic energy scale for a single band. 
For the case of two bands near or crossing the Fermi level, the
effective mass approximation is even more questionable.
Since in the bi-layer superconductors Bi 2:2:1:2
and Y 1:2:3 the band splitting of the even and odd bands is on the order
of tens of meVs, the approximation misses large terms corresponding to
interband transitions at relatively small energies.
Detailed knowledge of the magnitude of the scattering amplitude in this
case requires knowledge of the wave functions of the two states, which has
currently not been investigated. Thereby, use of the
effective mass approximation is uncontrolled in many band systems and 
comparisons to experiment intensities must be viewed with caution. A
discussion of this point is elaborated in the Appendix.
 
However, even if the above assumption does not hold in the actual experiments,
this gives an insight into the relative orders of $\gamma_{11}=\gamma_{22}$
and $\gamma_{1,2}=\gamma_{2,1}$, or in other words measures the splitting
between $\gamma_{+}$ and $\gamma_{-}$ (see Eq. (6)). Without interlayer
coupling the degenerate bands have the same dispersion such that
$\gamma_{11}({\bf k})=\gamma_{22}({\bf k})$, which holds generally
due to reflection symmetry.

An alternative approach is based on the experimental observation that
the spectra near optimal doping for a
wide range of cuprate materials depends only mildly on the incoming laser
frequency. Since the polarization orientations transform as various
elements of the point group of the crystal, one can use symmetry to classify
the scattering amplitude, viz.,
\begin{equation}
\gamma({\bf k};\omega_{I},\omega_{S})=\sum_{L}\gamma_{L}(\omega_{I},\omega_{S})
\Phi_{L}({\bf k}),
\end{equation}
where $\Phi_{L}({\bf k})$ are either Brillouin zone (B.Z.H., orthogonal 
over the entire Brillouin zone) or Fermi surface (F.S.H., orthogonal
on the Fermi surface only) harmonics which transform
according to point group transformations of the crystal\cite{allen}. 
Representing the magnitude but not the 
${\bf k}$-dependence of both intra- and interband scattering, the prefactors
can be approximated to be frequency independent and taken as model constants 
to fit absolute intensities. 
Thus we have simplified the many-band problem in terms
of symmetry components which can be related to charge degrees of freedom
on portions of the Fermi sheets. While sacrificing information
pertaining to overall intensities, we have gained the ability to probe
and compare excitations on different regions of the Fermi surface based
solely on symmetry classifications. This can be illustrated
by considering the various experimentally accessible polarization orientations.

Using an $x,y$ coordinate system locked to the CuO$_{2}$ planes, incident and 
scattered light polarizations aligned
along $\hat x+\hat y, \hat x-\hat y$ for example
transform according to $B_{1g}$ symmetry, and thus 
\begin{equation}
\Phi_{B_{1g}}({\bf k})=\cos(k_{x}a)-\cos(k_{y}a) + \dots,
\end{equation}
where $\dots$ are higher order B.Z.H.
Likewise, ${\bf e}^{I,S}$ aligned
along $\hat x, \hat y$ transforms as $B_{2g}$:
\begin{equation}
\Phi_{B_{2g}}({\bf k})= \sin(k_{x}a)\sin(k_{y}a) + \dots.
\end{equation}
The $A_{1g}$ basis function is
\begin{eqnarray}
&&\Phi_{A_{1g}}({\bf k})=
a_{0} + a_{2}[\cos(k_{x}a)+\cos(k_{y}a)] \\
&&+ a_{4}\cos(k_{x}a)\cos(k_{y}a)
+a_{6}[\cos(2k_{x}a)+\cos(2k_{y}a)] +\dots, \nonumber
\end{eqnarray}
where the expansion parameters $a_{i}$ determined via a fitting procedure
with experiment\cite{efm}. The $A_{1g}$ response is not directly accessible
from experiments and must be obtained by subtracting several combinations
of the response for various polarization orientations. 

By considering the ${\bf k}-$dependence of the basis functions, it is clear
that the $B_{1g}$ part of the spectra essentially probes light scattering
events along the $k_{x}$ or $k_{y}$ axes, $B_{2g}$ probes the diagonals,
and $A_{1g}$ is more of an average over the entire Brillouin zone. It is in 
this manner that information about the momentum dependence of the 
superconducting energy gap can be usefully extracted from the data\cite{one}.
That information lies in two important aspects: the low frequency power-
law behavior of the spectra and the positions of the low energy peaks.

At this point we can derive an expression for the cross section of the
two band system. First we define the ${\bf k}-$dependent Tsuneto function
$\lambda({\bf k},i\omega)$ as\cite{one,carbotte}
\begin{eqnarray}
&&\lambda({\bf k}, i\omega)=
{\Delta({\bf k})^{2}\over{E({\bf k})^2}}\\
&\times &\tanh\biggl[{E({\bf k})\over{2 T}}\biggr]
\left[{1\over{2E({\bf k})+i\omega}}+
{1\over{2E({\bf k})-i\omega}}\right].\nonumber
\end{eqnarray}
Here $E({\bf k})^{2}=\epsilon({\bf k})^{2}
+\Delta({\bf k})^{2}$ and we have set $k_{B}=\hbar=1$. 
We have neglected 
vertex corrections of the pairing interactions which while responsible for 
maintaining gauge invariance and producing collective modes\cite{kandd},
can be shown
to have only a limited effect on the spectra at low energies for $d-$wave
superconductors\cite{one}.

The Raman cross section is related to the Raman density correlation
function via the fluctuation-dissipation theorem,
\begin{equation}
I(\omega)\sim(1+n(\omega))\chi^{\prime\prime}(\omega),
\end{equation}
with 
\begin{equation}
\chi(i\omega)=\int_{0}^{1/T} d\tau e^{-i\omega\tau}
\langle T_{\tau}[\tilde\rho(\tau)\tilde\rho]\rangle,
\end{equation}  
with $T_{\tau}$ the time-ordering operator and the imaginary part is
obtained by analytic continuation, $i\omega \rightarrow \omega +i0$. 
In order to take into account the long-range Coulomb screening in the
limit of $q \rightarrow 0$, the following formula must be calculated
\begin{equation}
\chi_{sc}^{\prime\prime}(\omega)=\chi^{\prime\prime}(\omega)
-\left\{ {\chi_{\gamma,1}(\omega)\chi_{1,\gamma}(\omega)
\over{\chi_{1,1}(\omega)}}\right\}^{\prime\prime},
\end{equation}
where $\chi$ is given by Eq. (14), and
\begin{equation}
\chi_{\gamma,1}(i\omega)=\int_{0}^{1/T} d\tau e^{-i\omega\tau}\langle
T_{\tau}[\rho(\tau)\tilde\rho]\rangle=\chi_{1,\gamma}(i\omega);
\end{equation}
\begin{equation}
\chi_{1,1}(i\omega)=\int_{0}^{1/T} d\tau e^{-i\omega\tau}\langle
T_{\tau}[\rho(\tau)\rho]\rangle.
\end{equation}

It is straightforward
now to calculate the light scattering cross section for the two band
model using Eqs. (5-6) and (12-17). We arrive at the following compact
expression for the cross section:
\begin{eqnarray}
\chi^{\prime\prime}_{sc}(\omega)=
\sum_{{\bf k},\pm}[\gamma({\bf k})\pm\gamma_{1,2}({\bf k})]^{2}
\lambda_{\pm}^{\prime\prime}({\bf k},\omega) 
\\-
\left\{{\biggl( \sum_{{\bf k},\pm}[\gamma({\bf k})\pm\gamma_{1,2}({\bf k})]
\lambda_{\pm}({\bf k},\omega)\biggr)^{2}\over{\sum_{{\bf k},\pm}
\lambda_{\pm}({\bf k},\omega)}}\right\}^{\prime\prime}, \nonumber
\end{eqnarray}
where $\lambda_{\pm}$ is the Tsuneto function for the bonding/anti-bonding
band, with energy gap $\Delta_{\pm}({\bf k})$.
Rearranging terms, we can cast the result in terms of an addition
of the result for each single band plus a mixing term:
\begin{equation}
\chi^{\prime\prime}_{sc}(\omega)=\chi^{\prime\prime}_{+}(\omega)
+\chi^{\prime\prime}_{-}(\omega)+\Delta\chi^{\prime\prime}(\omega),
\end{equation}
where
\begin{eqnarray}
&&\chi^{\prime\prime}_{\pm}(\omega)=\\
&&\sum_{\bf k}\gamma_{\pm}^{2}({\bf k})
\lambda_{\pm}^{\prime\prime}({\bf k},\omega) 
-\left\{{\biggl(\sum_{\bf k}\gamma_{\pm}({\bf k})\lambda_{\pm}({\bf k},\omega)\biggr)^{2}
\over{\sum_{\bf k}\lambda_{\pm}({\bf k},\omega)}}\right\}^{\prime\prime},
\nonumber 
\end{eqnarray}
\begin{eqnarray}
&&\Delta\chi^{\prime\prime}(\omega)=\biggl\{{\bigl[
\sum_{\bf k}\lambda_{+}({\bf k},\omega)\bigr]\bigl[\sum_{\bf k}\lambda_{-}
({\bf k},\omega)\bigr]\over{\sum_{{\bf k},\pm}
\lambda_{\pm}({\bf k},\omega)}} \\
&\times& \biggl[{\sum_{\bf k}\gamma_{+}({\bf k})
\lambda_{+}({\bf k},\omega)\over{\sum_{\bf k}\lambda_{+}({\bf k},\omega)}}
-{\sum_{\bf k}\gamma_{-}({\bf k})
\lambda_{-}({\bf k},\omega)\over{\sum_{\bf k}\lambda_{-}({\bf k},\omega)}}
\biggr]^{2}\biggr\}^{\prime\prime}. \nonumber
\end{eqnarray}

Eqs. (19-21) are our central results. It shows that the total response can be
considered as a sum of the single band contributions in addition to
a mixing term which corresponds to odd combinations of fluctuations
on the bands simultaneously (see Fig. 1). In order to discern the
features of the mixing term, it is useful to write $\gamma_{\pm}({\bf k})
=\gamma_{\pm}^{0}+\Delta\gamma_{\pm}({\bf k})$, where the average of
$\Delta\gamma({\bf k})$ around the Fermi surface vanishes. The Fermi
surface average of a quantity $A({\bf k})$ is defined as
\begin{equation}
\langle A({\bf k}) \rangle_{F.S.} = {\sum_{\bf k}\delta(\epsilon({\bf k})-
\mu) A({\bf k})\over{\sum_{\bf k}\delta(\epsilon({\bf k})-\mu)}}.
\end{equation}

Then it useful to consider the following cases: 

(1) If the layers are uncoupled, then $\gamma_{+}({\bf k})=\gamma_{-}
({\bf k})$ and the mixing term vanishes: $\Delta\chi^{\prime\prime}
(\omega)=0$.

(2) If the scattering in the $A_{1g}$ symmetry channel $\gamma$ is
independent of ${\bf k}$ for each band (as in the case of
scattering on real charge), then
$\Delta\chi \sim (\gamma_{+}^{0}-\gamma_{-}^{0})^{2} \sim \gamma_{1,2}^{2}$.
This means that the light scattering induces an interlayer charge transfer.

(3) Considering when the light induces charge transfers on different
parts of the Fermi surfaces, $\Delta\gamma_{\pm}({\bf k})$ must be
included and includes different symmetry channels depending on the
orientation of the incident and scattered light polarizations. $A_{1g}$
symmetry reflects the approximate tetragonal symmetry of the CuO$_{2}$
planes, while the part of $\gamma$ showing symmetries different than
$A_{1g}$ (e.g., $B_{1g}, B_{2g}$ and $E_{g}$) changes sign when
symmetry transformations are applied. The Tsuneto function depends only
on the energy gap squared, and thus all the sums in Eq. (21)
{\it are zero except for the $A_{1g}$ component of $\gamma_{\pm}
({\bf k})$}. For tetragonal materials, this means that the $B_{1g},
B_{2g},$ and $E_{g}$ channels do not have a mixing term and therefore
the total response for these channels is simply the additive
contributions from each single band. The mixing term only contributes
to the $A_{1g}$ channel, the strength of which is roughly determined
by the interlayer coupling. 

In the following section, we will investigate Eqs. (19-21) in detail 
for the single layer compound with different Fermi surfaces (section IIIA.) and 
bi-layer Y 1:2:3 (section IIIB).

\section{evaluation for model superconductors and discussion of results}

We start by considering previous approaches\cite{one,carbotte,irwin,candk}. 
The majority of the calculations have been performed by restricting the
k-sum onto the Fermi surface\cite{one,irwin,candk}. While giving 
qualitatively insightful
description of the resulting Raman spectra, this approximation does not
capture important physics of charge dynamics over the entire Brillouin zone, 
such as the role of van Hove singularities, and since the energy gap is a 
sizeable portion of the bandwidth, neglecting light scattering which
breaks Cooper pairs at energies off the Fermi surface
will miss large contributions to the overall cross section. 
This has been pointed out in the recent calculations of Branch and Carbotte
\cite{carbotte}. Moreover, since
the $A_{1g}$ channel measures weighted averages over the Fermi surface, 
restricting the calculation for the $A_{1g}$ spectra to only the Fermi
surface would generally overemphasis the effect of small
changes of the Fermi surface manifold due to the presence of the gap
singularities. Therefore, in our opinion the entire k-sum must be performed.

In the following we confine ourselves to tetragonal systems with in-plane
lattice constant $a$ which can be modeled in the absence of interlayer coupling 
by the following band structure:
\begin{eqnarray}
&&\epsilon({\bf k})=-2t[\cos(k_{x}a)+\cos(k_{y}a)]+4t^{\prime}
\cos(k_{x}a)\cos(k_{y}a)\nonumber\\
&&-2t^{\prime\prime}[\cos(2k_{x}a)+\cos(2k_{y}a)]-\mu.
\end{eqnarray}
This form of the band structure results from a downfolding of an 8 band
Hamiltonian into a single band as described in Ref. \cite{oka}. The parameters
$t,t^{\prime}$, and $t^{\prime\prime}$ can either be taken from LDA 
calculations\cite{oka}
or be taken as parameters to fit the Fermi surfaces observed via
photoemission\cite{fit,photo}.

\subsection{Single layer case $t_{\perp}=0$}

In this section the $T << T_{c}$ Raman response for a single CuO layer
is obtained by evaluating the ${\bf k}$-sums directly for the
parameters for La 2:1:4\cite{levin} and Y 1:2:3\cite{oka}. 

The $B_{1g}$ and $B_{2g}$ responses are plotted in Fig. 3 using an energy
gap of $d_{x^{2}-y^{2}}$ symmetry, $\Delta({\bf k})=\Delta_{0}[\cos(k_{x}a)-\cos(k_{y}a)]/2$,
with $\Delta_{0}$ chosen to be 12 (30) meV for La 2:1:4 (Y 1:2:3), respectively, in order
to match the position of the $B_{1g}$ peak to that seen in experiment\cite{exp}. 
In both cases, the $B_{1g}$ response for low
frequencies varies as $\omega^{3}$ while the $B_{2g}$ varies linearly
with $\omega$. This can be shown to be generally true for these channels
for any number of higher harmonics used, and results from a
consideration of the density of states (DOS), which varies 
linearly with energy due to the nodes, and the behavior of the Raman 
vertex $\gamma$ near the 
gap nodes. Since the $B_{2g}$ vertex is finite near the nodes, the DOS 
determines the low frequency behavior. However, the gap and the vertex vanish 
at the same place for the $B_{1g}$ channel, which in turn leads to the cubic 
behavior. This delicate interplay of vertex and energy gap can thus provide
a unique determination of the nodal gap behavior.

Moreover, additional information lies in the peaks of the spectra. A smooth
peak is seen for $B_{2g}$ and a double-peak is obtained for $B_{1g}$. The
lower peak is due to the energy gap while the other comes from the van
Hove singularity. The peak of the spectra due to superconductivity in both
cases is higher for the 
$B_{1g}$ channel ($\omega_{peak}=2\Delta_{max}$)
than for $B_{2g}$, ($\omega_{peak}\sim 1.7\Delta_{max}$) since
the gap maxima are located at the
same part of the Fermi surface where the $B_{1g}$ basis function is
largest, and where the $B_{2g}$ basis function vanishes. This is generally
true regardless of Fermi surface shape, although the relative positions of
the peaks can be affected slightly by changing the underlying manifold 
(see Figs. 3a and 3b and Refs. \cite{one,carbotte,irwin}). 
The shape of the spectra can also be slightly modified 
by considering higher
harmonics of the energy gap \cite{carbotte} and/or by including final-state
interactions\cite{one} or impurity scattering\cite{imp}.
Apart from the last case, the power-law behaviors at low frequencies
are unaffected and therefore their observation is a robust check on the
symmetry of the energy gap. 

The van Hove peak only shows up in the $B_{1g}$ channel since the 
$B_{2g}$ channel assigns no weight to the location of the van Hove points. 
The van Hove peak lies at
a higher energy determined by its location off the Fermi surface and the
value of the energy gap there. There are indications
that a double-peak feature has been seen in underdoped Y 1:2:3
with a T$_{c}=60 K$\cite{slakey}. As the van Hove moves further away from the
Fermi surface for example with doping, the peak shifts to higher 
frequencies and has a smaller
residue. Inelastic scattering at higher energies will also act to smear
out this feature and thus a single smeared and perhaps asymmetric peak
would be seen if the van Hove lies quite near the Fermi surface, and could
in principle move the $B_{1g}$ peak slightly above $2\Delta_{max}$. This
may be reflected in the experiments of Ref. \cite{alt} which saw a
sensitivity of the $B_{1g}$ peak position to doping.

In Fig. 4 we show a calculation of
the unscreened and screened $A_{1g}$ response for a 
$d_{x^{2}-y^{2}}$ paired superconductor for the Y 1:2:3 parameters. We
have chosen the coefficients $a_{i}$ in Eq. (11) using the effective mass
approximation\cite{efm}. By again considering the
nature of the nodal structure of the energy gap, since the $A_{1g}$ response
measures a weighted average of charge fluctuations around the Fermi surface,
it can be shown that the low energy part of the $A_{1g}$ response must vary 
with energy in the same way as the DOS. 

The screened and unscreened responses at higher energies are very different. 
For the unscreened response, two peaks arise 
corresponding to the singular contribution of pair breaking
along the axes and the van Hove contribution as in the $B_{1g}$ channel. 
Screening completely 
reorganizes the spectra to remove both peak contributions, replacing
them with a much smoother function with more spectral weight at lower 
energies and a much reduced spectral weight at higher energies. This is 
due to the screening function (2nd part of Eq. (15)) which also contains 
peaks at $2\Delta_{max}$ due to the weighting along the $k_{x}$ and $k_{y}$
axes, and the van Hove. These peaks cancel the unscreened peaks and create a 
mild shoulder lying near $2\Delta_{max}$, as seen in Fig. 4. 

These results are different than those obtained previously for the case of
a simpler evaluation of Eq. (18) on a cylindrical Fermi surface
\cite{one} and for a more thorough evaluation of Eq. (18) using the full
k-sum and a consistent treatment of the harmonics for both the band
structure, Raman vertices, and energy gap\cite{carbotte}. Since publication 
of Ref. \cite{one}, a numerical error was detected in
the code which once corrected also produced curves similar to the one
shown in Fig. 4. This is treated and corrected in a forthcoming publication
\cite{tpddeerr}. We also note that Ref. \cite{candk} also produced
an $A_{1g}$ response which is peaked near $2\Delta$, but remark that these
calculations were performed by neglecting the real parts of Eq. (20).

In an effort to understand these differences we have investigated Eq. (20) 
for various choices of band structure, Raman vertices, and additional 
harmonics added to the energy gap. While we found that the linear power-law 
rise of the $A_{1g}$ response is robust to changes in parameters, the 
position of the peak in the spectra can vary anywhere less than or
equal to $2\Delta$ depending on the choice of parameter sets.
This suggests that the calculations for the $A_{1g}$ screened response are 
unstable to the inclusion of higher order Brillouin zone harmonics for 
either the energy gap, band structure, or the Raman vertex. 

Specifically, we have included the next
few harmonics for a $d_{x^{2}-y^{2}}$ energy gap, 
\begin{eqnarray}
&&\Delta({\bf k})=\Delta_{0}\biggl\{[\cos(k_{x}a)-\cos(k_{y}a)]/2 +
\\
&&{\Delta_{1}\over{8}}[\cos(k_{x}a)-\cos(k_{y}a)]^{3}+
{\Delta_{2}\over{32}}[\cos(k_{x}a)-
\cos(k_{y}a)]^{5}\biggr\},\nonumber
\end{eqnarray}
and used the parameters $a_{i}$ in Eq. (11) for the vertex derived
from the band structure\cite{efm}. The results are
summarized in Fig. 5, which shows the various forms for the $A_{1g}$
screened response using different combinations of energy gap
harmonics. This result suggest that the weighting
of the $A_{1g}$ vertex in the Brillouin zone depends strongly on a
variety of parameters and can easily weight regions where the energy
gap is quite different for small changes in the parameters. This results
in varying positions of the peak. In general, including higher harmonics
for the energy gap pushes the peak in the $A_{1g}$ channel downward
from near $2\Delta_{max}$, which is in agreement with the results 
obtained by Branch and Carbotte\cite{carbotte}. We are then to conclude
that the position of the $A_{1g}$ peak can not provide useful information
without detailed knowledge of the fine structure of the energy gap itself
even for the case of a single band near the Fermi level.

In this manner, similar symmetry considerations can be made for various 
types of energy gaps. The Raman spectra calculated for various energy gaps
is summarized in Table 1 and compared to the low energy DOS $N(\omega)$.
The dominant contribution to the Raman lineshape
is due to the location and behavior of the energy gap near both the nodes and 
the maxima. Subsequently, different energy gaps produce different line
shapes as summarized in Table 1. Here it is important to note that the
cubic rise of the spectra in any channel requires that both the vertex
and the energy gap have the same behavior near the nodes. In 
tetragonal systems, this requires the presence of a $d_{x^{2}-y^{2}}$ energy 
gap. By considering small orthorhombic distortions, the $A_{1g}$ and $B_{1g}$ 
channels become
mixed and therefore a linear rise with frequency could be obtained at low
energies in a region determined by the amount of symmetry breaking.
Since experimentally
the low frequency part of the $B_{1g}$ spectra rises cubically in the most
tetragonal systems, while in Y 1:2:3 the spectra may have an intrinsic small 
linear part,
this strongly suggests the likelihood of a $d_{x^{2}-y^{2}}$ pair state 
for these systems. Exploration of the low
frequency part of the spectra could put stringent constraints on other
pair-state candidates.

\subsection{Bi-Layer}

In this section we consider the bi-layer superconductor Y 1:2:3.  
For the interlayer coupling $t_{\perp}$, we follow Ref. \cite{oka} and define
\begin{equation}
t_{\perp}({\bf k})=t_{\perp}[\cos(k_{x}a)-\cos(k_{y}a)]^{2},
\end{equation}
with $t_{\perp}/t=0.2$. In the absence of inter-layer coupling the chemical
potential is chosen so that $\langle n \rangle=0.8$.
The two Fermi surface sheets are shown in Fig. 6 for these choices of the 
parameters. Lastly, we choose the energy gaps on the two bands to be of the
$d_{x^{2}-y^{2}}$ type, $\Delta_{\pm}({\bf k})=\Delta_{\pm}
[\cos(k_{x}a)-\cos(k_{y}a)]/2$, and further choose $\Delta_{\pm}=30$ meV for 
both bands for simplicity. 

The resulting spectra for the bonding and anti-bonding bands, as well as
the mixing term for $A_{1g}$ are shown in Fig. 7. Here again we have used
the effective mass approximation for the expansion parameters of Eq. (11)
for illustration\cite{efm}, although as indicated earlier this approximation
does not hold for this bi-layer system. Considering the $B_{1g}$ and $B_{2g}$
channels, which are simply additive, it can be inferred from experimental
result which shows only one peak in these channels that either the
energy gap is nearly identical both in symmetry and magnitude or alternatively
that the Raman response is given predominantly by one band. Otherwise
two distinct peaks would appear (if gaps have different symmetry and/or
magnitude) and identical power-law behavior for each channel
would be seen (if the gaps
were of different symmetry but same magnitude)\cite{cave}. Since the band
splitting is relatively small in Y 1:2:3 and even smaller in Bi 2:2:1:2,
the second possibility in our opinion is unlikely\cite{photo}. 
Therefore, the energy
gap on each band must be similar on energy scales determined by the amount
of inelastic scattering which would smear a double-peak feature. Again we
note that the sharp nature of the $B_{1g}$ peak would imply that its
position and shape would be most sensitive to the location of the van Hove
points for each band and the maximal value of the energy gaps\cite{slakey,alt}.
In fact, as shown in the Fig. 8, the van Hove peak in some cases can be stronger than
the $2\Delta_{max}$ peak which could erroneously lead to an overestimated
value of $2\Delta_{max}/$T$_{c}$.

Due to the small bi-layer splitting the mixing term for $A_{1g}$ is small, 
as shown in Fig. 7. Two very small peaks with opposite curvature
at roughly $2\Delta_{max}$ for each of the bands are seen due 
to the odd combination of interlayer charge transfer, as suggested by 
Krantz and Cardona. These peaks are non-divergent and are strongly
suppressed compared to the contributions from the individual bands.
Since adding electron correlations
suppresses the bi-layer splitting even further\cite{oka2,photo}
the mixing term is only of minor importance and
the resulting spectra can be well approximated as the sum of the 
contributions from the two individual bands. In this way, a consistent
picture can emerge since experimentally the electronic Raman spectra do
not differ substantially in the bi-layer or single-layer materials
\cite{one,exp}.

We have also checked the effect of abandoning the effective mass approximation
for the vertex as well as adding additional harmonics to the energy gap,
as considered in the previous section. Our results are shown in Fig. 8
for the parameters chosen to most closely model the experimental data
\cite{exp}:
$$ B_{1g}: \gamma({\bf k}) \sim \cos(k_{x}a)-\cos(k_{y}a),$$
$$ B_{2g}: \gamma({\bf k}) \sim \sin(k_{x}a)\sin(k_{y}a)-
[\sin(k_{x}a)\sin(k_{y}a)]^{3},$$
$$ A_{1g}: \gamma({\bf k}) \sim a_{4}\cos(k_{x}a)\cos(k_{y}a)+$$
$$a_{6}[\cos(2k_{x}a)+\cos(2k_{y}a)]; a_{4}=-2t^{\prime}-t_{\perp}, 
a_{6}=4t^{\prime\prime}+t_{\perp},$$
$$ \Delta_{2}=0.5, \Delta_{0}=30 {\rm meV}.$$
The results show that the peak of the $A_{1g}$ spectra moves downward in
frequency with the inclusion of the additional energy gap harmonic as for
the single band case, while the mixing term and power-law
behavior remains unaffected. The $B_{2g}$ peak is pushed to lower
frequencies by the inclusion of the higher harmonic for the energy gap
as seen in Ref. \cite{carbotte}, but can be shifted back by including the
next harmonic for the Raman vertex. The resulting peak
lies roughly 30 percent lower than the peak in the $B_{1g}$
spectrum. The van Hove feature does not change appreciably and thus the 
$B_{1g}$ channel changes are quite small. The power-law behavior is
unaffected, as remarked earlier. From these results we can then conclude
that the Raman response for the bi-layer can be well modelled by the
added response calculated for the two separate energy bands with the only
difference is that the use of the effective mass approximation is invalid
(see the Appendix for more details).

\section{Conclusions}

We have seen that electronic Raman scattering contains a wealth of information
concerning the nature of superconductivity in unconventional superconductors
which can be extracted from symmetry considerations alone. 
When applied to the experimental data for ${\it optimally}$ doped cuprate 
materials,
good agreement can be obtained for $d_{x^{2}-y^{2}}$ pairing in single band 
materials, and nearly identical gaps of $d_{x^{2}-y^{2}}$ symmetry 
for bi-layer systems. The analysis is based on the prominent behavior of
the spectra at low frequencies and the approximate position of the 
maxima for each channel. It is also worthwhile to point out that 
similar considerations can be applied to the optically active phonon lineshapes
below $T_{c}$\cite{phonons}. 

In this paper we confined ourselves to the simple case of clean, weakly
coupled unconventional superconductivity. As pointed out earlier, like
other correlation functions Raman in clean systems is insensitive to the
phase of the order parameter around the Fermi surface. Moreover, as a
consequence of taking the $q \rightarrow 0$ limit, the considered spectra
vanishes as one approaches T$_{c}$ due to phase space constraints. Both
of these deficiencies have been remedied recently by considering both
the effect of impurity scattering\cite{imp,forth}. As shown previously, 
disorder effects on unconventional
as opposed to conventional superconductors can in principle be used to 
determine whether the gap averages to zero around the Fermi surface\cite{bork}.
In particular, the disorder dependence of the exponent of the low frequency 
and/or 
temperature dependence of the $B_{1g}$ channel provided a useful signature of 
gap phase properties\cite{imp,forth}. While the resulting spectra could be 
applied
to low frequencies, the flat ``Marginal'' behavior at high frequencies cannot
be reproduced by this method. However, that can be corrected by considering
spin fluctuation scattering or electron-electron scattering due to 
nesting\cite{imp,forth,imp2}.
However, more work is needed to correctly describe the full channel dependent
Raman spectra for large frequency ranges for optimally doped cuprates.

Further, we have seen that the lineshape of the $A_{1g}$ screened response 
is sensitive to parameter changes and to the number of harmonics used
for the energy gap. However, the low frequency behavior and the general
``flatness'' of the lineshape seems to be independent of parameter
choices. We remark that a consistent treatment of the number
of harmonics used for the band structure, Raman vertices, and energy gap
must be considered to give meaningful results for the peak position. 
Without a more detailed knowledge of the {\it actual} ${\bf k}$-dependence 
of the energy gap (symmetry is not enough), a theoretical description of the 
peak of the spectrum in the $A_{1g}$ channel is incomplete. In addition,  a
recent study found that the position of the $A_{1g}$ peak 
did not move under pressure, in contrast to the peak in the $B_{1g}$
channel\cite{syass}, which may indicate that the peak in $A_{1g}$ may
not be directly related to the charge degrees of freedom. Clearly, further 
work is needed.

Moreover, further work is needed in order to understand the properties of
the Raman spectra {\it away} from optimal doping\cite{exp,slakey,alt,pri}. 
It is suggested that the $B_{1g}$ peak may be the most sensitive to doping
due to (1) the sharpness of the peak, (2) the role of the van Hove singularity,
and (3) the sensitivity of the Fermi surface along the $k_{x},k_{y}$ axes 
to small changes of the band structure parameters. However, in order to 
understand the effects of doping and to arrive at a more complete picture 
of Raman scattering in the cuprates, a particular model of
superconductivity is needed as is a more sophisticated approach for
handling the spin degrees of freedom.

\acknowledgements
We wish to thank M. V. Klein, J. C. Irwin, R. Hackl, D. Einzel, M. Cardona, 
J. Carbotte, D. Branch, and G. Blumberg for many enlightening discussions.
This work was supported by the U.S.-Hungarian Science and Technology
Joint Fund under Project Number 265, NSF Grant No. DMR 95-28535, and by
the Hungarian National Research Fund under Grants No. OTKA T016740,
T021228, 7283, and T020030. One of the authors (T.P.D.)
would like to acknowledge the hospitality of the Institute of Physics
at the Technical University of Budapest, the Research Institute for
Solid State Physics, and the Walther Meissner Institute f\"ur
Tieftemperaturforschung, where parts of this work were completed.

\section{Appendix: Extension to $N$ bands crossing the Fermi level}

In this appendix we develop formulas for electronic Raman scattering
applicable for a multiband model. Our goal is to show that

{\it i)} the use of the effective mass approximation for the Raman
vertex is unjustified in case of more than one band crossing the Fermi
level, and

{\it ii)} there can be no divergence in the $A_{1g}$ Raman spectrum
in the superconducting state.

First we recall that the differential electronic Raman scattering
cross section is given by\cite{kandd}
\begin{equation}
{d^2\sigma\over d\Omega d\omega}={\hbar\over\pi}r_0^2{\omega^s\over
\omega^i}\left [ 1-e^{-\hbar\omega/k_BT}\right ]^{-1}{\rm Im}
\chi_{\tilde\rho^+,\tilde\rho}(\omega),
\end{equation}
where $\omega=\omega^i-\omega^s$ is the difference between the
incoming and scattered light frequency, $r_0=e^2/mc^2$ is the Thomson
radius, and
\begin{equation}
\chi_{\tilde\rho^+,\tilde\rho}(t)=
{i\over\hbar}\Theta(t)\langle [\tilde\rho^+({\bf q},t),
\tilde\rho({\bf q},0)]\rangle.
\end{equation}
For a Bloch electron system with one-electron
states $|n,{\bf k}\rangle $ the effective density operator is
represented as
\begin{equation}
\tilde\rho({\bf q})=\sum_{n,n^\prime}\sum_{{\bf k},\sigma}
\gamma_{n,n^\prime}({\bf k})c^+_{n,{\bf k},\sigma}c_{n^\prime,{\bf k}
-{\bf q},\sigma},
\end{equation}
with ${\bf q}={\bf q}^i-{\bf q}^s$ is the difference of the incoming
and scattered light wavenumber. In the experimental situation relevant
for Raman scattering $q^i,q^s\ll k_F$, and the Raman vertex simplifies
to\cite{ag}
\begin{eqnarray}
&&\gamma_{n,n^\prime}({\bf k})={\bf e}^i{\bf e}^s\delta_{n,n^\prime}\\
&&+{1\over m} 
\sum_\nu\biggl[ {\langle n,{\bf k}|{\bf e}^s{\bf p}|\nu,
{\bf k}\rangle\langle\nu,{\bf k}|{\bf e}^i{\bf p}|n^\prime,{\bf k}
\rangle\over\varepsilon_{n^\prime}({\bf k})-\varepsilon_\nu({\bf k})
+\hbar\omega^i}\nonumber \\
&&+{\langle n,{\bf k}|{\bf e}^i{\bf p}|\nu,{\bf k}\rangle
\langle\nu,{\bf k}|{\bf e}^s{\bf p}|n^\prime,{\bf k}\rangle\over
\varepsilon_{n^\prime}({\bf k})-\varepsilon_\nu({\bf k})-\hbar\omega^s}
\biggr],\nonumber
\end{eqnarray}
where ${\bf e}^i$ and ${\bf e}^s$ are the polarization vectors of the
incoming and scattered light, ${\bf p}=-i\hbar\nabla$, and
$\varepsilon_n({\bf k})$ is the energy of the Bloch electron. The
correlation function of the effective density is now readily
evaluated as
\begin{eqnarray}
&&\chi_{\tilde\rho^+,\tilde\rho}(\omega)=\sum_{n,n^\prime}\sum_{{\bf k},
\sigma}|\gamma_{n,n^\prime}({\bf k})|^2\\
&&\times{f[\varepsilon_n({\bf k})]-
f[\varepsilon_{n^\prime}({\bf k}-{\bf q})]\over\hbar\omega+i\delta
-\varepsilon_n({\bf k})+\varepsilon_{n^\prime}({\bf k}-{\bf q})}.\nonumber
\end{eqnarray}
The above susceptibility consists of both intraband ($n=n^\prime$) and
interband ($n\neq n^\prime$) contributions.

In the context of an $N$ band model it is implicitly assumed that out
of the infinite bands of a Bloch system only $N$ bands (with typical
bandwidth $W$) cross the Fermi level, while all the other bands are
far above or below those $N$ bands of interest, separated by a gap
$G$ of order $G\gg W$. This is the largest energy scale of the problem,
therefore when we consider Raman scattering in this model we must
assume $\hbar\omega^i,\hbar\omega^s\ll G$, although the energies of the
incoming and scattered light may well exceed those of the optical
transitions between bands close to the Fermi level (non resonant
scattering). Under these assumptions only terms with indices $n$ and
$n^\prime$ both labeling bands crossing the Fermi level will
contribute to the Raman susceptibility in Eq.(30). The relevant Raman
vertex will further simplify to
\begin{eqnarray}
&&\gamma_{n,n^\prime}({\bf k})={\bf e}^i{\bf e}^s\delta_{n,n^\prime}
+{1\over m}\times\\
&&\sum_{\nu}^{\prime}\biggl[\langle n,{\bf k}|{\bf e}^s{\bf p}|\nu,{\bf k}
\rangle\langle\nu,{\bf k}|{\bf e}^i{\bf p}|n^\prime,{\bf k}\rangle +\nonumber \\
&&\langle n,{\bf k}|{\bf e}^i{\bf p}|\nu,{\bf k}\rangle\langle\nu,{\bf k}|
{\bf e}^s{\bf p}|n^\prime,{\bf k}\rangle\biggr][
\varepsilon_{n^\prime}({\bf k})-\varepsilon_\nu({\bf k})]^{-1},\nonumber
\end{eqnarray}
where the prime on the summation indicates that $\nu$ runs only over
indices belonging to the faraway bands. Since $n$ and $n^\prime$ belong
to bands close to the Fermi level, the light energies are negligible
compared to the large band gaps in the denominators of Eq. (30). On the
other hand when $\nu$ assumes values belonging to the "close" bands,
the band gaps are negligible compared to the light energies, and it is
easily seen, that those terms yield contributions of the order of
$W\hbar\omega/\hbar\omega^i\hbar\omega^s\approx \omega/\omega^{i,s}$.
Under usual experimental
circumstances ($\omega\ll\omega^i$) these terms can be neglected in the
Raman vertex.

Since we are interested in the low frequency response and interband
transitions contribute to Im$\chi_{\tilde\rho^+,\tilde\rho}$ only
above some finite threshold, in the following we restrict ourselves to
intraband ($n=n^\prime$) transitions. The Raman susceptibility is now
given by
\begin{equation}
\chi_{\tilde\rho^+,\tilde\rho}(\omega)=\sum_{n,{\bf k},\sigma}[
\gamma_n({\bf k})]^2{f[\varepsilon_n({\bf k})]-f[\varepsilon_n({\bf k}-
{\bf q})]\over\hbar\omega+i\delta-\varepsilon_n({\bf k})+
\varepsilon_n({\bf k}-{\bf q})},
\end{equation}
where the Raman vertex is
\begin{equation}
\gamma_n({\bf k})={\bf e}^i{\bf e}^s+{1\over m}\sum_\nu ^\prime{\langle
n,{\bf k}|{\bf e}^s{\bf p}|\nu,{\bf k}\rangle\langle\nu,{\bf k}|{\bf e}^i
{\bf p}|n,{\bf k}\rangle+c.c.\over\varepsilon_n({\bf k})-
\varepsilon_\nu({\bf k})}
\end{equation}
real, and an even function of ${\bf k}$. The susceptibility is additive
for those bands crossing the Fermi level, however the Raman vertex reduces
to the familiar effective mass formula\cite{ag}
\begin{equation}
\gamma_n({\bf k})=\sum_{\alpha,\beta}e^i_\alpha{m\over\hbar^2}{\partial^2
\varepsilon_n({\bf k})\over\partial k_\alpha\partial k_\beta}e^s_\beta
\end{equation}
only for the case of a single band crossing the Fermi level. This is because
in the effective mass formula\cite{AM}
$\nu$ runs over all values except $n$, while
for more than one bands crossing the Fermi level there are some other terms
also excluded from the sum for the Raman vertex in Eq.(33). The missing
terms are in general of order unity due to the relatively small gaps
between bands close to the Fermi level. In the Raman vertex however,
these terms are negligible because of the large laser light energies
(see Eq.(29)).
Therefore in a multiband problem knowledge of
the electronic energies $\varepsilon_n({\bf k})$ is not sufficient for a
precise evaluation of the Raman vertices $\gamma_n({\bf k})$. Using the
effective mass formula in this situation renders the analysis qualitative
at best.

Consider our system in the superconducting state. Assuming that
superconductivity develops independently in the bands crossing the Fermi
level, the intraband Raman response will again be additive, and the
well known formula\cite{kandd} is generalized as
\begin{equation}
\chi_{\tilde\rho,\tilde\rho}(\omega)=\sum_{n,{\bf k}}[\gamma_n({\bf k})
]^2\lambda_n({\bf k},\omega),
\end{equation}
where
\begin{equation}
\lambda_n({\bf k},\omega)=\tanh\left [{E_n({\bf k})\over 2k_BT}\right ]{
4|\Delta_n({\bf k})|^2/E_n({\bf k})\over 4E_n^2({\bf k})-(\hbar\omega+
i\delta)^2}
\end{equation}
is the Tsuneto function for the $n^{th}$ band involving the superconducting
order parameter $\Delta_n({\bf k})$ and quasiparticle energy $E_n^2({\bf k})
=\varepsilon_n^2({\bf k})+|\Delta_n({\bf k})|^2$
of that band. Although it is not trivial, it can be shown that 
for the $N$ band model outlined above the Raman
vertex derived for the normal state in Eq. (33) is an excellent approximation
for the one to be used in the superconducting state\cite{Vup}.

The screening effect of the long range Coulomb interaction can be taken
into account by including coupling of the effective density to the
density ($\rho$) fluctuations\cite{KF}. In the long wavelength limit
appropriate for Raman scattering the screened response is given 
by\cite{one,kandd}
\begin{equation}
\chi_{\tilde\rho,\tilde\rho}^{scr}=\chi_{\tilde\rho,\tilde\rho}-
{\chi_{\tilde\rho,\rho}\chi_{\rho,\tilde\rho}\over\chi_{\rho,\rho}},
\end{equation}
with
\begin{equation}
\chi_{\tilde\rho,\rho}(\omega)=\sum_{n,{\bf k}}\gamma_n({\bf k})
\lambda_n({\bf k},\omega)=\chi_{\rho,\tilde\rho}(\omega),
\end{equation}
and
\begin{equation}
\chi_{\rho,\rho}(\omega)=\sum_{n,{\bf k}}\lambda_n({\bf k},\omega).
\end{equation}
Note that there is no interband contribution to $\chi_{\tilde\rho,
\rho}$ and $\chi_{\rho,\rho}$ in the long wavelength limit
appropriate for the present analysis. 
On the other hand, the second term in Eq. (37) mixes the bands
(unless $\chi_{\tilde\rho,\rho}=0$), thereby making
the screened response non additive. The
screening effect of the long range Coulomb interaction can be made
more transparent by rewriting Eq. (37) with the help of Eqs. (35),
(38) and (39) as
\begin{equation}
\chi_{\tilde\rho,\tilde\rho}^{scr}(\omega)=\sum_{n,{\bf k}}[
\gamma_n({\bf k})-\bar\gamma(\omega)]^2\lambda_n({\bf k},\omega),
\end{equation}
where the average of the Raman vertex (which is screened out) involves
all the relevant bands and depends on frequency as
\begin{equation}
\bar\gamma(\omega)={\sum_{n,{\bf k}}\gamma_n({\bf k})\lambda_n({\bf k},
\omega)\over\sum_{n,{\bf k}}\lambda_n({\bf k},\omega)}.
\end{equation}

The Raman response as given by Eqs. (40) and (41) has been thoroughly
investigated for the case of only one band crossing the Fermi
level\cite{one}. It was shown, that for tetragonal symmetry and for a
quasi two-dimensional electron system only the $A_{1g}$ response is
screened. Moreover, since the Tsuneto function picks up divergent
contributions only for $\omega=2|\Delta|_{max}$, and those contributions
come from the neighborhood of points on the Fermi surface where the
maximum value of
$|\Delta|$ is attained, the necessary condition for divergent Raman
response is that the Raman vertex does not vanish at those points.
For example, in case of a superconducting gap of $d_{x^2-y^2}$
symmetry, the $B_{1g}$ response is divergent at $\omega=2|\Delta|_{max}$
translating into a robust peak at that frequency in experiments, while
there is no divergence (anywhere) in the $B_{2g}$ response. Therefore
the frequency of any observed maximum in that spectrum is not related to
the superconducting gap maximum. The unscreened $A_{1g}$ response is
also divergent at $2|\Delta|_{max}$, but this divergence is removed by
screening, since $\bar\gamma$ at $\omega=2|\Delta|_{max}$ takes exactly
the same value as $\gamma({\bf k})$ does at the critical ${\bf k}$ points.
This is because for $\omega=2|\Delta|_{max}$ the Tsuneto function
assigns infinite weight in $\bar\gamma$ to exactly that $\gamma$ value.
Therefore again, any peak in the experimental $A_{1g}$ spectrum is
not directly related to the superconducting gap.

In a multiband situation the question might arise as to whether the above
one band result holds for the removal of any divergences in the $A_{1g}$
symmetry. Fueling doubts about it is the fact that in a one band model
a momentum independent Raman vertex would completely be screened out
leading to no response at all, while in the multiband case different but
still momentum independent $\gamma_n$ values will obviously lead to
finite response according to Eq. (40). The existence of a divergence
however is a more subtle problem. Any divergence can only come from one
of the Tsuneto functions at one of the different frequencies
$\omega_n=2|\Delta_n|_{max}$. There is no reason to assume that
the maximum gap is the same for any two bands if those bands are
different. For a given $\omega_n$ only $\lambda_n$ can give rise to
divergence, the contribution of all the other bands in Eq. (40)
is finite at that frequency. Similarly, $\bar\gamma(\omega_n)$ will be
determined solely by the divergent contributions from the $n^{th}$
band both in the numerator and denominator of Eq.(41). Therefore as far
as the existence of divergences in the $A_{1g}$ spectrum is concerned,
the multiband problem reduces to the one band problem with the same
result: any possible divergence at the various gap maxima
in the Raman spectrum is removed by screening.

\onecolumn
\begin{table}
\centerline{TABLE I}
\centerline{Summary of Raman response for various pair state candidates for 
clean, tetragonal superconductors.$^{*}$} 
\begin{tabular}{|l|l|l l|l l|l l|} \hline
{$\Delta({\bf k})$} & {$N(\omega \rightarrow 0)$} &
\multicolumn{2}{c|}{$B_{1g}$}   &
\multicolumn{2}{c|}{$B_{2g}$}   & \multicolumn{2}{c|}{$A_{1g}$}  
\\ 
& & {$\chi^{\prime\prime}(\omega \rightarrow 0)$} & ${\omega_{peak}\over{\Delta_{max}}}$ & {$\chi^{\
prime\prime}(\omega \rightarrow 0)$} & ${\omega_{peak}\over{\Delta_{max}}}$ & 
{$\chi^{\prime\prime}(\omega \rightarrow 0)$} & 
${\omega_{peak}\over{\Delta_{max}}}^{\dagger\dagger\dagger}$\\
\hline
isotropic $s-$wave & $\Theta(\omega-\Delta)$& $\Theta(\omega-2\Delta)$ & 2 & 
$\Theta(\omega-2\Delta)$ & 2 & 
$\Theta(\omega-2\Delta)$& 2\\  \hline
$d_{x^{2}-y^{2}}$ & $\omega$ & $\omega^{3}$ & 2 & $\omega$ & $\sim 1.7$ & $\omega$ &  variable  
\\ \hline
$s+id_{x^{2}-y^{2}}$ & $\Theta(\omega-\Delta_{s})$ & 
$\Theta(\omega-2\Delta_{s})$ & 2 & 
$\Theta(\omega-2\Delta_{s})$ & $2\Delta_{s}/\Delta_{d}$ & $\Theta(\omega-2\Delta_{s})$ & $2\Delta_{s}/\Delta_{d} $ 
\\ \hline
$d_{xy}$ & $\omega$ & $\omega$ & $\sim 1.7$ & $\omega^{3}$ & 2 & $\omega$ & 
 variable 
\\ \hline
$(d_{x^{2}-y^{2}})^{m}$ & $\omega^{1/m}$ & $\omega^{3/m}$ & 2 & $\omega^{1/m}$ & $< 1.7$ & $\omega^{1/m}$ & variable
\\ \hline
$g-$wave$^{\dagger}$ & & & & & & & \\
$\sim \cos(k_{x}a)\cos(k_{y}a)$ & $\omega$ & $\omega$ & 2 & $\omega$ & 0.6 & $\omega$ & variable
\\ \hline
extended $s-$wave$^{\dagger\dagger}$ & & & & & & & \\
$\sim \cos(k_{x}a)+\cos(k_{y}a)$ & $\omega$ & $\omega$ & $\sim 1.7$ & $\omega$ & 2 & $\omega$ & variable
\\ \hline \hline
experimental results & & & & & & & \\
for optimal doping & $\omega$ & $\omega^{3}$ & 2 & $\omega$ & $1.5\sim 1.8$ & 
$\omega$ & $1.0 \sim 1.2 $
\\ \hline
\end{tabular}
$^{*}$ Here phase is undetermined and thus $\mid d_{x^{2}-y^{2}}\mid$ would
yield the same as $d_{x^{2}-y^{2}}$. Moreover, orthorhombic distortions 
will mix $B_{1g}$ and $A_{1g}$ channels.\\ 
$^{\dagger}$ Parameters chosen to split node at $45^{o}$ into two 
modes at $45^{o} \pm 5^{o}$ and a subsidiary 
maximum equal to 20 percent of gap maxima. \\
$^{\dagger\dagger}$ Parameters chosen such that $\langle n \rangle=0.8$.\\
$^{\dagger\dagger\dagger}$ $A_{1g}$ peak position depends sensitively on
harmonic representation of band structure, vertices, and energy gap (see
text for a discussion).\\
\end{table}
\newpage
\begin{figure}[t]
\epsfxsize=3.5in
\epsfysize=4.0in
\epsffile{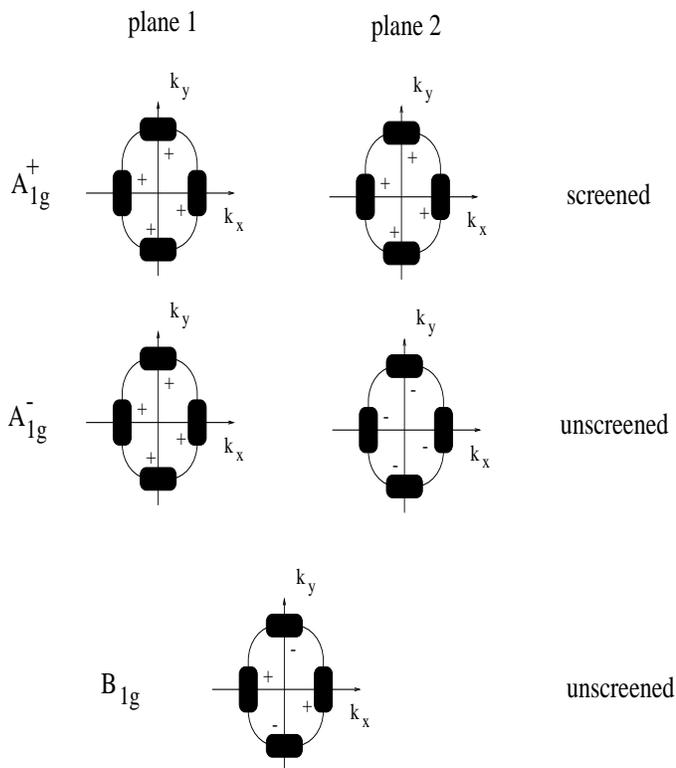}
\caption{Schematic picture of Raman density weighting for $A_{1g}^{\pm}$ and
$B_{1g}$ channels for bi-layer systems. For nondegenerate bands, the
$-$ combination for the $A_{1g}$ channel can remain unscreened.}
\end{figure}
\begin{figure}[t]
\epsfxsize=3.5in
\epsfysize=4.0in
\epsffile{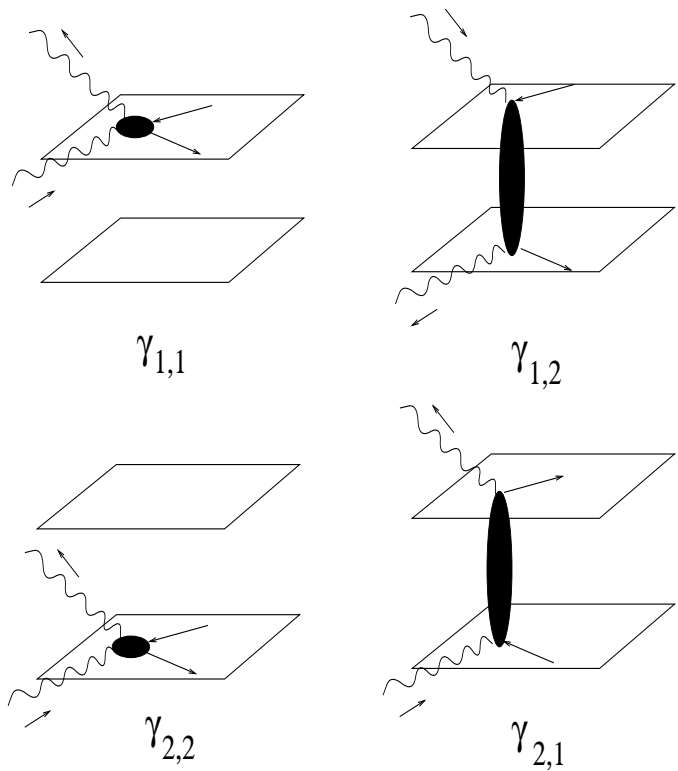}
\caption{Schematic picture for the various sorts of light scattering for
a bi-layer system.}
\end{figure}
\newpage
\begin{figure}[t]
\epsfxsize=2.5in
\epsfysize=4.0in
\epsffile{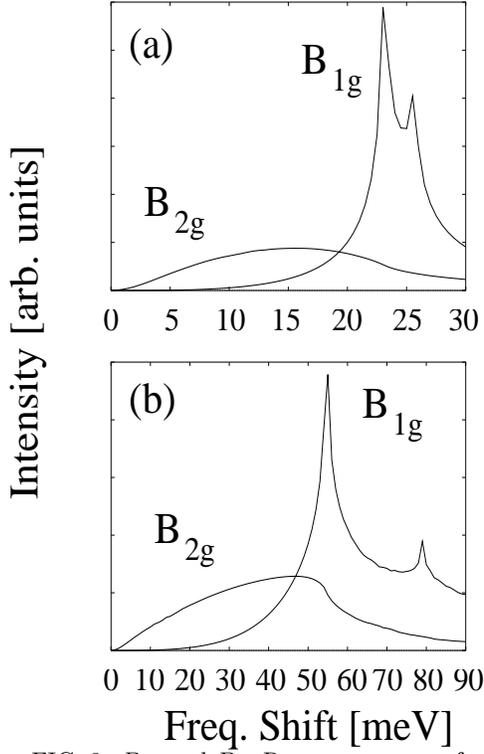}
\caption{$B_{1g}$ and $B_{2g}$ Raman responses for a single band 
plotted for parameters sets for (a) La 2:1:4 and (b) Y 1:2:3.
Here a filling $\langle n \rangle =0.8$ and $t=100$ meV, and 
$\Delta(k)=\Delta_{0}[\cos(k_{x}a)-\cos(k_{y}a)]/2$ are used for both figures. 
The other parameters are: (a) $t^{\prime}/t=0.16, t^{\prime\prime}=0, 
\Delta_{0}=12$meV; (b) $t^{\prime}/t=0.2, 
t^{\prime\prime}/t=0.25, \Delta_{0}=30$ meV. The small peak at higher frequencies
in the $B_{1g}$ channel is due to the van Hove singularity.} 
\end{figure}
\begin{figure}[t]
\epsfxsize=2.5in
\epsfysize=3.5in
\epsffile{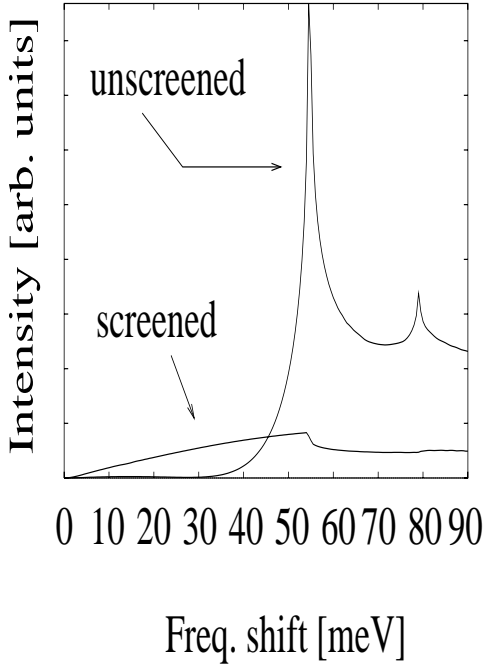}
\caption{$A_{1g}$ response calculated for a single band 
with and without Coulomb screening for the Y 1:2:3 parameters.}
\end{figure}
\newpage
\begin{figure}[t]
\vskip -1.0cm
\epsfxsize=2.5in
\epsfysize=2.5in
\epsffile{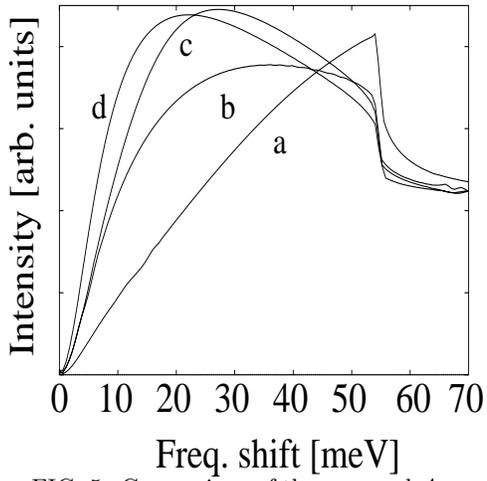}
\caption{Comparison of the screened $A_{1g}$ response obtained for
different number of gap harmonics used than for Fig. 4 (which is redrawn
here and labelled as (a)). All parameters used are the same as Fig. 4
except the following:
(a) no changes, (b) $\Delta_{0}=32$ meV, $\Delta_{1}=1$, (c) $\Delta_{0}=
35$ meV, $\Delta_{2}=1$ (d) $\Delta_{0}=
36$ meV, $\Delta_{1}=\Delta_{2}=1$.}
\end{figure}
\begin{figure}[t]
\epsfxsize=2.5in
\epsfysize=2.5in
\epsffile{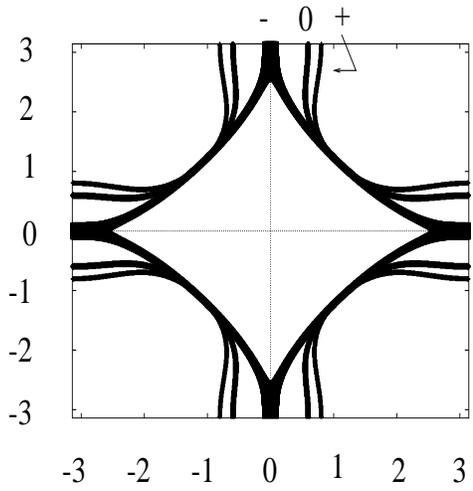}
\vskip 1.0cm
\caption{Fermi surface for the absence (labelled as 0) and presence of 
interlayer coupling (labeled $\pm$ for the bonding/anti-bonding bands).}
\end{figure}
\newpage
\begin{figure}[t]
\vskip 5.0cm
\epsfxsize=2.5in
\epsfysize=3.5in
\epsffile{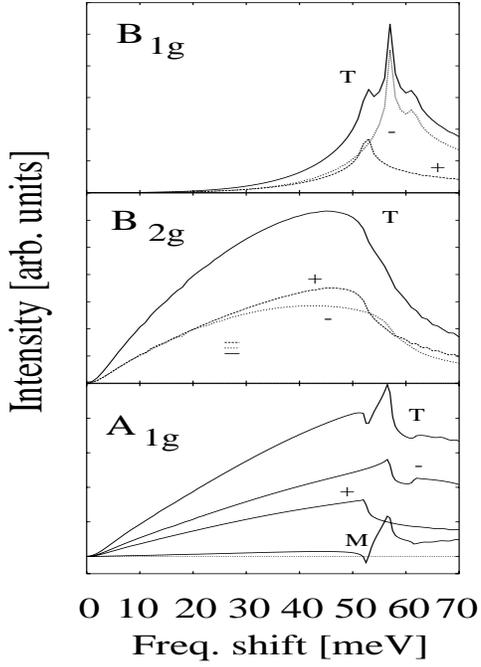}
\vskip 0.5cm
\caption{Raman response calculated for Y 1:2:3 bi-layer for various channels
as indicated. The $+ (-)$ indicates the bonding (anti-bonding) 
band, respectively, the symbol $M$ is the mixing term [Eq.
(21) which only contributes for the $A_{1g}$ channel], and the total response
is indicated by the symbol $T$. The peaks are located a different positions for
the $B_{1g}$ channel due to the different energy dispersions of the two
bands. The van Hove peak lies at much higher energies for the bonding band due to
its larger separation from the Fermi level.}
\end{figure}
\newpage
\begin{figure}[t]
\vskip 5.0cm
\epsfxsize=2.5in
\epsfysize=3.5in
\epsffile{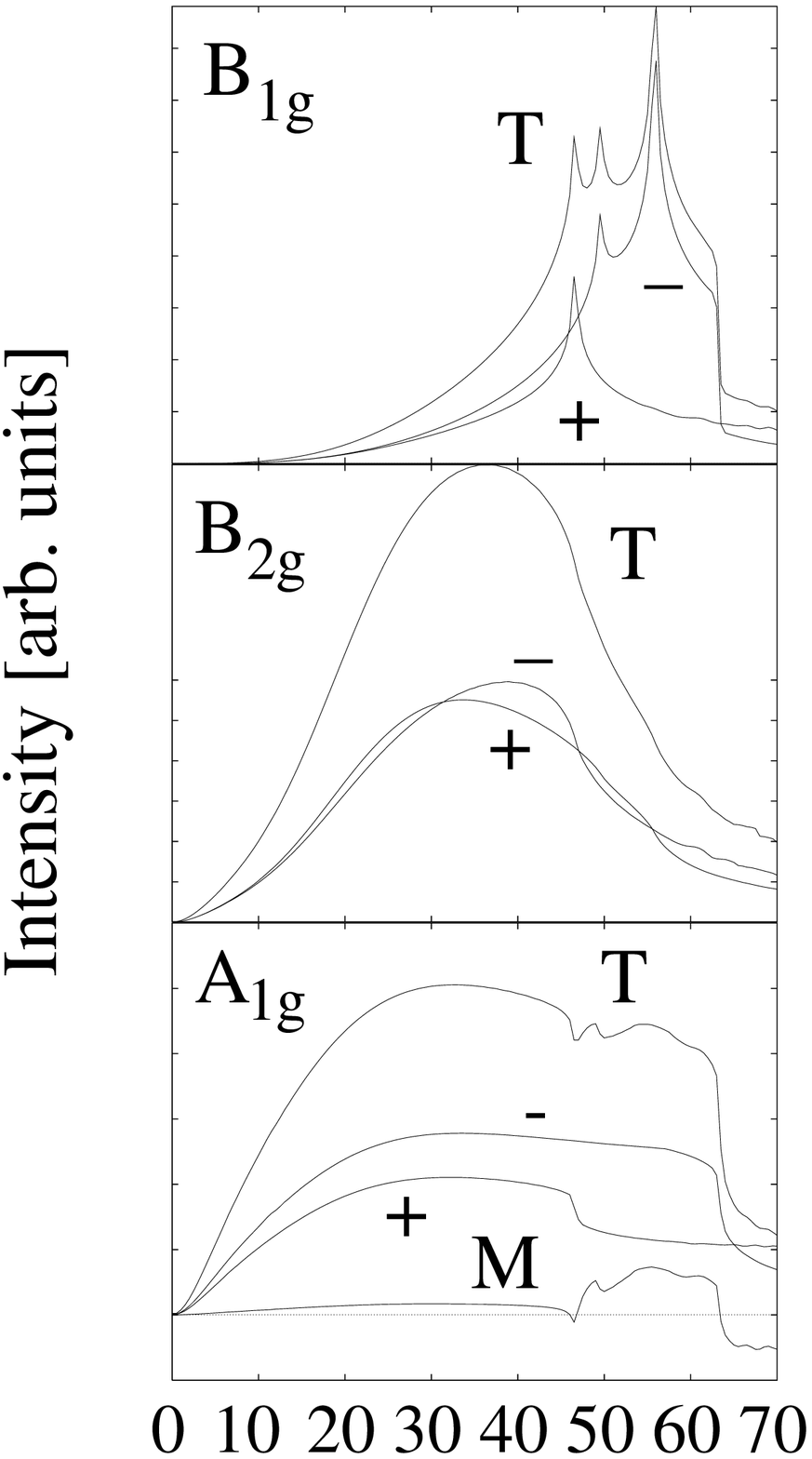}
\vskip 0.5cm
\caption{Raman response calculated for Y 1:2:3 bi-layer for various channels
as indicated for different parameter sets than those considered in
Fig. 7. See text for details.}
\end{figure}

\begin{references}
\bibitem{one}
T. P. Devereaux {\it et al.}, Phys. Rev. Lett. {\bf 72}, 396 (1994);
Phys. Rev. B {\bf 51}, 16336 (1995); Journ. of Supercond. {\bf 8}, 421 (1995);
Journ. of Phys. Chem. Solids {\bf 56}, 1711 (1995). 
\bibitem{carbotte}
D. Branch and J. P. Carbotte, Phys. Rev. B {\bf 52}, 603 (1995);
to appear in Phys. Rev. B.
\bibitem{irwin}
X. K. Chen {\it et al.}, Physica C {\bf 227}, 113 (1994);
{\it ibid.}, 1089 (1995) [These calculations did
not include Coulomb interactions].
\bibitem{candk}
M. C. Krantz and M. Cardona, Phys. Rev. Lett. {\bf 72}, 3290
(1994); Journ. of Low Temp. Phys. {\bf 99}, 205 (1995).
\bibitem{exp}
S. L. Cooper {\it et al}, Phys. Rev. B {\bf 37}, 5920 (1988);
R. Hackl {\it et al., ibid} {\bf 38}, 7133 (1988);
S. L. Cooper {\it et al., ibid}, 11934 (1988); T. Staufer {\it et al.}, Phys. 
Rev. Lett. {\bf 68}, 1069 (1992); A. Yamanaka {\it et al.}, Phys. Rev. B
{\bf 46}, 516 (1992); R. Nemetschek {\it et al.}, Phys. Rev. B
{\bf 47}, 3450 (1993); X. K. Chen {\it et al.}, Phys. Rev. Lett. {\bf 73}, 3290 (1994); 
{\it ibid}. Journ. of Supercond. {\bf 8}, 495 (1995);
A. Hoffmann {\it et al.}, Physica C {\bf 235-240}, 1897 (1994);
C. Kendziora {\it et al}, Phys. Rev. B {\bf 52}, 9867 (1995), and
references therein.
\bibitem{slakey}
F. Slakey {\it et al.}, Phys. Rev. B {\bf 42}, 2643 (1990).
\bibitem{alt}
X. K. Chen {\it et al.}, Phys. Rev. B {\bf 48}, 10530 (1993).
\bibitem{semin}
G. Contreras, A. K. Sood and M. Cardona, Phys. Rev. B {\bf 32},
924 (1985);
I. P. Ipatova, A. V. Subashiev and V. A. Voitenko, Sol. State
Commun. {\bf 37}, 893 (1981).
\bibitem{photo}
Z. X. Shen {\it et al.}, Physics Rep. {\bf 253}, 1 (1995);
Science {\bf  267} 343 (1995); M. R. Norman {\it et al.}, Phys. Rev. 
B {\bf 52} 15107 (1995); H. Ding {\it et al.}, Phys. Rev. Lett. {\bf 76},
1533 (1996).
\bibitem{kandd}
M. V. Klein and S. B. Dierker, Phys. Rev. B {\bf 29}, 4976 (1984);
H. Monien and A. Zawadowski, Phys. Rev. B {\bf 41}, 8798 (1990).
\bibitem{ag}
A. A. Abrikosov and V. M. Genkin, Zh. Eksp. Teor. Fiz. {\bf 40}, 842 (1973) 
[Sov. Phys. JETP {\bf 38}, 417 (1974)].
\bibitem{allen}
P. Allen, Phys. Rev. B {\bf 13}, 1416 (1976).
\bibitem{efm}
In the effective mass approximation Eq. (11) yields 
$a_{0}=0, a_{2}=2t, a_{4}=-8t^{\prime}$ and 
$a_{6}=8t^{\prime\prime}$.
\bibitem{levin}
Q. Si {\it et al.}, Phys. Rev. B {\bf 47}, 9055 (1993).
\bibitem{oka}
O. K. Andersen et. al., Phys. Rev. B {\bf 49}, 4145 (1993); Journ. of Phys.
Chem. Solids {\bf 56}, 1573 (1995). 
\bibitem{fit}
R. J. Radtke {\it et al.}, preprint;
G. Blumberg {\it et al.}, Phys. Rev. B {\bf 52}, 15741 (1995).
\bibitem{imp}
T. P. Devereaux, Phys. Rev. Lett. {\bf 74}, 4313 (1995).
\bibitem{cave}
The power-law behavior would not be changed if one of the gaps has the 
same magnitude as the other but completely isotropic $s$-wave. However,
then all channels would show a sharply defined peak at $2\Delta_{max}$.
This is in disagreement with experiment and appears to be unlikely. 
\bibitem{oka2}
A. I. Liechtenstein {\it et al.}, LANL preprint 9509101; M. Langer {\it et al.},
Phys. Rev. Lett. {\bf 75}, 4508 (1995).
\bibitem{tpddeerr}
T. P. Devereaux and D. Einzel, erratum to be published.
\bibitem{phonons}
T. P. Devereaux {\it et al.}, Phys. Rev. B {\bf 51}, 505 (1995), and 
references therein.
\bibitem{forth}
T. P. Devereaux, to be published in Int. Journ. of Mod. Phys. B.
\bibitem{bork}
L. S. Borkowski and P. J. Hirschfeld, Phys. Rev. B {\bf 49}, 15404 (1994), 
and references therein.
\bibitem{imp2}
C. Jiang and J. P. Carbotte, Sol. State Commun. {\bf 95}, 643 (1995).
\bibitem{syass}
A. F. Goncharov {\it et. al}, AIP Conference Proceedings {\bf 309}, 707 (1994);
preprint.
\bibitem{pri}
R. Hackl, J. C. Irwin, C. Kendziora and M. V. Klein, private communication.
\bibitem{AM} N. W. Ashcroft and N. D. Mermin, {\it Solid State
Physics} (Saunders College, Philadelphia, 1976), p. 766.
\bibitem{Vup} A. Virosztek (unpublished).
\bibitem{KF} L. P. Kadanoff and I. I. Falko, Phys. Rev. {\bf 136},
1170 (1964).
\end{references}
\end{document}